\documentclass[twocolumn]{emulateapj}

\usepackage{graphicx}
\usepackage{epsfig}

\newcommand{\clusterage}{2.4}         
\newcommand{\nummembers}{480}  
\newcommand{\precision}{0.4}          
\newcommand{\thatbinid}{003002}       
\newcommand{\clustrvmean}{2.3}    
\newcommand{\fancyclustcent}{2.34}
\newcommand{\fancyclusterr}{0.05}   
\newcommand{\clustrvdev}{1.0}     

\newcommand{\expcontam}{33}    
\newcommand{\expsingmem}{364}  
\newcommand{\perccontam}{ 9\%}  

\newcommand{\numcatU}{130}  
\newcommand{\numcatSM}{397}  
\newcommand{\numcatSN}{475}  
\newcommand{\numcatBM}{41}  
\newcommand{\numcatBN}{11}  
\newcommand{\numcatBLM}{42}  
\newcommand{\numcatBLN}{69}  
\newcommand{\numcatBU}{42}  

 \newcommand{\numprimestars}{1454}  
 \newcommand{\numprimeobserved}{1207}  
 \newcommand{\numprimefinal}{924}  
 \newcommand{\percprimeobserved}{83\%}  
 \newcommand{\percprimefinal}{64\%}  

 \newcommand{\numprimebright}{436}
 
\newcommand{\numdbobserved}{1207}  
\newcommand{\numdbmeasurements}{6571}  

\newcommand{\numwiynobserved}{1102}  
\newcommand{\numwiynmeasurements}{5455}  

\newcommand{\numcfaall}{191}  
\newcommand{\numcfaobserved}{170}  
\newcommand{\numcfameasurements}{733}  
\newcommand{\numcfaprime}{170}  

\newcommand{\numbinsolns}{52}  
\newcommand{\numbinnosolns}{153}  
\newcommand{\numbinall}{205}  
\newcommand{\numvarmem}{83}  

\newcommand{\numprimebinnosolns}{\numbinnosolns}
\newcommand{\numsingmembers}{\numcatSM}

\newcommand{\numbinCOMmem}{\numcatBM}

\newcommand{\WIYNfoot}{The WIYN Observatory is a joint facility of the
University of Wisconsin-Madison, Indiana University, Yale University,
and the National Optical Astronomy Observatories.}

\newcommand{\MMTfoot}{The Multiple Mirror Telescope is a joint facility
of the Smithsonian Institution and the University of Arizona.}

\shorttitle{WOCS: Radial Velocities of NGC~6819}
\shortauthors{Hole et al.}

\begin{document}

\title{WIYN Open Cluster Study. XXIV.
	Stellar Radial-Velocity Measurements in NGC~6819}

\author{K. Tabetha Hole\altaffilmark{1,2} (kth@astro.wisc.edu)} 
\author{Aaron M. Geller\altaffilmark{1,2} (geller@astro.wisc.edu)}
\author{Robert D. Mathieu\altaffilmark{1,2} (mathieu@astro.wisc.edu)}
\author{Imants Platais\altaffilmark{3} (imants@pha.jhu.edu)}
\author{S{\o}ren Meibom\altaffilmark{1,2,4} (smeibom@cfa.harvard.edu)}
\author{David W. Latham\altaffilmark{4} (dlatham@cfa.harvard.edu)}

\altaffiltext{1}{Astronomy Department, U. Wisconsin-Madison, 475 N
Charter St, Madison WI 53706}
\altaffiltext{2}{Visiting Astronomer, Kitt Peak National Observatory, 
National Optical Astronomy Observatory, which is operated by the 
Association of Universities for Research in Astronomy (AURA) under 
cooperative agreement with the National Science Foundation.}
\altaffiltext{3}{Department of Physics and Astronomy, Johns Hopkins
University, 3400 North Charles Street, Baltimore MD 21218}
\altaffiltext{4}{Harvard-Smithsonian Center for Astrophysics, 60 Garden
Street, Cambridge MA 02138}

\begin{abstract}

We present the current results from our ongoing radial-velocity survey
of the intermediate-age (\clusterage\ Gyr) open cluster NGC 6819.  Using
both newly observed and other available photometry and astrometry we
define a primary target sample of \numprimestars\ stars that includes
main-sequence, subgiant, giant, and blue straggler stars, spanning a
magnitude range of 11$\leq$$V$$\leq$16.5 and an approximate mass range
of $1.1$ to $1.6\ M_{\sun}$. Our sample covers a 23 arcminute (13 pc)
square field of view centered on the cluster.
We have measured \numdbmeasurements\ radial velocities for an unbiased
sample of \numdbobserved\ stars in the direction of the open cluster
NGC~6819, with a single-measurement precision of $\precision\
\mathrm{km\ s^{-1}}$ for most narrow-lined stars.  We use our
radial-velocity data to calculate membership probabilities for stars
with $\geq$ 3 measurements, providing the first comprehensive membership
study of the cluster core that includes stars from the giant branch
through the upper main sequence.  We identify \nummembers\ cluster
members. Additionally, we identify velocity-variable systems, all of
which are likely hard binaries that dynamically power the cluster.
Using our single cluster members, we find a cluster average radial velocity of
\fancyclustcent\ $\pm$ \fancyclusterr\ $\mathrm{km\ s^{-1}}$.  We use
our kinematic cluster members to construct a cleaned color-magnitude
diagram from which we identify rich giant, subgiant, and blue straggler
populations and a well-defined red clump.  The cluster displays a 
morphology near the cluster turnoff clearly indicative of core
convective overshoot.  Finally, we discuss a few stars of note, one of
which is a short-period red-clump binary that we suggest may be the
product of a dynamical encounter.

\end{abstract}


\section{Introduction}

Intermediate-age open clusters (1-5 Gyr), like NGC 6819, provide critical
tests for theories of stellar evolution, as these clusters generally
display signs of convective core overshoot (i.e., ``blue hook''
morphologies) at the main-sequence turnoffs.  This distinctive
morphology is believed to be caused by the rapid contraction of
hydrogen-depleted convective cores in stars with masses $M \gtrsim 1.2\
M_{\sun}$.  At the edges of these cores there is a complex interplay
between radiative and hydro-dynamical processes such that the convective
cells can `overshoot' the classical core edge and mix material to
regions outside of the core. The detailed structures of main-sequence
turnoffs provide
readily available tests of theoretical models of stellar evolution which
include convective core overshooting \citep[e.g.,][hereafter
RV98]{rosvick98}.  Defining the turnoff structure requires excellent
photometry, no contamination from field stars, and removal of confusion
from the composite light of binaries \citep[e.g.,][]{daniel94}.

Additionally, studies of open clusters can reveal how stellar dynamics
influences pathways in stellar evolution. Blue stragglers are the
best-known example, but detailed studies of the 4 Gyr cluster M67
and the 7 Gyr cluster NGC 188 have revealed stars with a variety of
non-standard evolutionary paths, including products of dynamical 
interactions, mass transfer, mergers, etc. \citep{mathieu86, vandenberg01,
mathieu03, sandquist03, geller08}.  Many of these stars and star systems
are likely the products of binary encounters leading to stellar
exchanges and mergers, and provide a rich array of alternative stellar
evolution paths. Again, maximum confidence in membership is required to
identify these stars.  Both, proper-motion and radial-velocity (RV)
membership studies are critical, as photometric
determinations of membership by their nature will usually exclude
cluster stars with non-standard evolutionary histories.

NGC 6819 has been moderately well studied, yet until now the cluster
has lacked a comprehensive kinematic membership study which includes
stars from the giant branch through the upper main sequence.  There have
been multiple photometric studies of NGC 6819 that have helped to define
the cluster color-magnitude diagram (CMD) \citep{burkhead71, lindorf72,
auner74,rosvick98,kalirai01}.  The most recent estimates for the cluster
parameters suggest an age between 2.4 and 2.5 Gyr, (M-m)$_V$=12.3, an
E(B-V) between 0.10 and 0.16, and [Fe/H]$\sim -0.05$.  \citet{kang02}
found evidence for mass segregation in their photometric study of the
cluster. \citet{street02,street03,street05} have discovered numerous
photometrically variable stars.  \citet{sanders72} performed the first
and only astrometric membership study, covering a circular area
($r=18\arcmin$) centered on the cluster, and calculated memberships for
189 stars down to $V\sim14.5\ \mathrm{mag}$ reaching the red
giants and blue stragglers.  \citet{glushkova93}, \citet{friel89} and
\citet{thogersen93} performed limited RV studies of the NGC 6819 field,
quoting cluster mean RVs that range from $+4.8\pm0.9$ km s$^{-1}$ to
$+1\pm6$ km s$^{-1}$.

The location of NGC6819 in Cygnus places it within the field of view
of the Kepler space mission - a search for transiting earth-like planets.
Over the planned 4 year mission, Kepler will provide time-series photometric
observations with a 30 minute cadence and parts-per-million precision to
$V\simeq$17. Observations by Kepler therefore offer a unique opportunity
to study phenomena of stellar photometric variability, and to do so for
even Gyr old stars for which such variability can be of very small amplitude.
Observations of members of NGC\,6819 with Kepler will make possible studies
of e.g. stellar rotation and asteroseismology, as well as searches for
extra-solar planets and eclipsing binaries, among 2.5\,Gyr old stars over
a range of masses and evolutionary stages. 

We present the first comprehensive high-precision RV survey of the core
of NGC 6819 as part of the WIYN Open Cluster Study
\citep[WOCS;][]{mathieu00}.  Our data cover stars from the giant
branch through the upper main sequence and include many potential blue
stragglers, thereby providing a valuable membership database and the
first census of the hard-binary\footnote{A hard binary is defined as having an internal energy that is much greater than the energy of the relative motion of a single star moving within the cluster \citep{heggie74}.  For solar mass stars in a cluster with a one-dimensional velocity dispersion equal to 1 km s$^{-1}$, all hard binaries have periods less than $\sim$10$^5$ days.} population.
First, we present our
analysis of all CCD photometry and astrometry of the cluster currently
available (\S~\ref{s:phot}), from which we define our stellar sample
(\S~\ref{s:samp}).  We then describe our RV observations, data reduction
and precision in \S~\ref{odp}.  For stars with $\geq$3 RV measurements,
we calculate RV membership probabilities and identify RV variable stars
(\S~\ref{s:results}).  Our data show that NGC~6819 is a rich cluster,
with \nummembers\ RV selected members in the cluster core with masses in
the range of 1.1 to $1.6\ M_{\sun}$.  Our RV measurements provide excellent 
membership discrimination, crucial because of the cluster's
location in the Galactic plane.  The cleaned CMD (discussed in
\S~\ref{disc}) reveals rich populations of giants, subgiants and blue
stragglers, and a morphology near the cluster turnoff
indicative of core convective overshoot.  We also identify a few stars
of note, including one short-period red-clump binary that we suggest may
be the result of a dynamical encounter.  Future papers will analyze the
dynamical state of the cluster (e.g., mass segregation and velocity
dispersion), and study the hard-binary fraction and frequency of orbital
parameters.

\section{Cluster Photometry and Coordinates}
\label{s:phot}

Two primary goals of the WOCS study of NGC~6819 are to provide
high-quality $UBVRI$ photometry and astrometry. For a preliminary report
on the WOCS photometric study see \citet{sarrazine03}. The astrometric
study is underway. Yet to begin our RV survey of the cluster, we
required such data in order to define our stellar sample. Thus we
conducted a critical analysis of all the photometry and astrometry
available to us for use in this paper, and provide the results of this
analysis here. We use the 2MASS survey\footnote{This publication makes
use of data products from the Two Micron All Sky Survey, which is a
joint project of the University of Massachusetts and the Infrared
Processing and Analysis Center/California Institute of Technology,
funded by the National Aeronautics and Space Administration and the
National Science Foundation.}
\citep{skrutskie06} as the backbone of this analysis,
and define a complete set of 6166 stars in the direction of NGC 6819
within the magnitude range of 11$\leq$$V$$\leq$16.5 and extending to 30
arcminutes from the cluster center. From this set we have selected our
stellar sample for the RV survey of the cluster (as explained in
\S~\ref{def_primary}).

\subsection{Photometry}

\begin{deluxetable*}{ccccccccccccccccccccc}
\tabletypesize{\tiny} 
\tablewidth{0pc} 
\tablecolumns{21}
\tablecaption{
The first ten lines of the NGC 6819 WOCS photometry database. 
See \S \ref{s:phot} for a description of the photometry sources. 
\label{t:phot}}
\tablehead{
\colhead{WOCS} &
\colhead{} &
\multicolumn{4}{c}{Phot98} &
\colhead{} &
\multicolumn{2}{c}{Phot03} &
\colhead{} &
\multicolumn{2}{c}{RV98} &
\colhead{} &
\multicolumn{2}{c}{K01} &
\colhead{} &
\multicolumn{2}{c}{2MASS} &
\colhead{} &
\colhead{Auner} &
\colhead{Sanders} 
\\
\cline{3-6}  
\cline{8-9}   
\cline{11-12}   
\cline{14-15}   
\cline{17-18}  
\\
\colhead{ID\tablenotemark{a} } &
\colhead{} &
\colhead{V} &
\colhead{B-V} &
\colhead{V-R} &
\colhead{V-I} &
\colhead{} &
\colhead{V} &
\colhead{B-V} &
\colhead{} &
\colhead{V} &
\colhead{B-V} &
\colhead{} &
\colhead{V} &
\colhead{B-V} &
\colhead{} &
\colhead{J} &
\colhead{J-K} &
\colhead{} &
\colhead{ID\tablenotemark{b} } &
\colhead{ID\tablenotemark{c} } 
}
\startdata
001001 &  &    \nodata & \nodata & \nodata & \nodata &  &    12.692 &   1.229 &  &    12.669 &   1.245 &  &   \nodata & \nodata &  &    10.212 &   0.929 &  &      974  &    115  \\ 
002001 &  &     13.357 &   1.248 &   0.622 &   1.250 &  &    13.399 &   1.223 &  &    13.378 &   1.238 &  &   \nodata & \nodata &  &    10.341 &   0.722 &  &      435  & \nodata \\ 
003001 &  &     13.622 &   1.194 &   0.591 &   1.194 &  &    13.664 &   1.143 &  &    13.644 &   1.164 &  &   \nodata & \nodata &  &    11.564 &   0.673 &  &      395  &    116  \\ 
004001 &  &     13.936 &   1.203 &   0.612 &   1.208 &  &    13.994 &   1.177 &  &    13.971 &   1.191 &  &    14.001 &   1.170 &  &    11.847 &   0.683 &  &      428  &    100  \\ 
011001 &  &     15.701 &   0.618 &   0.316 &   0.719 &  &    15.746 &   0.591 &  &    15.720 &   0.609 &  &    15.763 &   0.600 &  &    14.436 &   0.315 &  &      396  & \nodata \\ 
013001 &  &     15.853 &   0.679 &   0.392 &   0.797 &  &    15.886 &   0.689 &  &    15.873 &   0.684 &  &   \nodata & \nodata &  &    14.501 &   0.482 &  &      432  & \nodata \\ 
015001 &  &     15.933 &   0.621 &   0.334 &   0.713 &  &    15.992 &   0.607 &  &    15.883 &   0.619 &  &    15.969 &   0.600 &  &    14.603 &   0.378 &  &      429  & \nodata \\ 
001002 &  &    \nodata & \nodata & \nodata & \nodata &  &    11.733 &   1.615 &  &    11.837 &   1.489 &  &   \nodata & \nodata &  &     8.819 &   0.996 &  &      550  &    110  \\ 
002002 &  &    \nodata & \nodata & \nodata & \nodata &  &    12.662 &   0.379 &  &    12.645 &   0.361 &  &   \nodata & \nodata &  &    11.794 &   0.180 &  &      976  &    105  \\ 
003002 &  &    \nodata & \nodata & \nodata & \nodata &  &    12.760 &   1.111 &  &    12.763 &   1.130 &  &   \nodata & \nodata &  &    10.660 &   0.736 &  &      390  &    126  \\ 

\enddata
\renewcommand{\arraystretch}{1.00}
\tablenotetext{a}{See \S~\ref{s:wocsid} for an explanation of the WOCS ID number.}
\tablenotetext{b}{The ID number from \citet{auner74}, if available.}
\tablenotetext{c}{The ID number from \citet{sanders72}, if available.}
\end{deluxetable*}

There are four sources of CCD $BV$ photometry available to us for the
cluster: RV98, \citet{kalirai01} (hereafter K01), and two unpublished
sets taken by members of the WOCS collaboration. (We will use ``RV98''
and ``K01'' to refer both to the papers and to the photometry sets used
in each, depending on context.) The two sets of WOCS photometry were
obtained at what was the KPNO and is now the WIYN\footnote{\WIYNfoot}
0.9m telescope. In March 1998, C. Dolan and R. Mathieu took $BV$ CCD
images to begin this RV project, hereafter called Phot98. These
observations used the Tektronics CCD (T2KA) at f/7.5, centered on the
cluster at $\alpha=19^{\rm h}41^{\rm m}19\fs3$ (J2000)
$\delta=+40\arcdeg 11\arcmin$, covering a $23\arcmin$ square (13 pc at 2
kpc distance) field-of-view and a magnitude range of $12.7<V<22.0$.  The
WOCS $UBVRI$ study (hereafter Phot03) is based on images obtained with
the SITe S2KB CCD camera (pixel size 0\farcs60), yielding a FOV of
$20\arcmin$.

The reductions of the Phot03 CCD frames include the usual bias correction
and flat-fielding using sky flats. Standard stars were drawn from the
list of \citet{landolt92}.  Instrumental magnitudes were obtained using
the DAOPHOT aperture photometry package \citep{stetson87}. The
photometric calibration equations included zero-point, linear-color, and
extinction terms.  The Phot98 images were taken on a non-photometric
night which precluded applying an absolute calibration. Instrumental
magnitudes were calibrated using a zero-point and color-term derived
from stars in common between Phot03 and Phot98.

We choose to analyze the zero-points of all the $BV$ CCD photometries
against Phot03, in the sense `$\Delta=$target-Phot03'. In all cases
only the common stars with $V<18.5$ are used. By construction, the
$BV$ system of Phot98 is identical to Phot03. For RV98, we find a mean
$\Delta V=-0.002$ and mean $\Delta (\bv) = +0.009$. For K01 the
offsets are $\Delta V =+0.018$ and $\Delta (\bv) =+0.004$.  The formal
uncertainty of the means is 0.004 magnitudes. Considering the high
degree of uniformity between the various photometries, when combining
$BV$ photometries we choose to subtract the offset for only the K01
$V$-magnitudes. All four sets of $BV$ photometry are given in Table
\ref{t:phot}, along with 2MASS $JK$ photometry. (Other bands and cross
identifications to Sanders and Auner are also provided, where
available.) The final $BV$ photometry, shown with our RV measurements
in Table \ref{the-table}, is derived as the means of all available $V$
magnitudes and $(\bv)$ colors.  \footnote{If there were more than two
  sets of photometry for a given star, a discordant value was excluded
  if it exceeded by $3\sigma$ the estimated scatter at a given
  magnitude.}

We note that our $BV$ photometry covers a $\sim$28 arcminute square
field of view centered on the cluster center (effectively, the spatial
extent of the K01 photometric study) and contains 2724 stars.  In 2005,
we chose to add additional stars observed by 2MASS to extend our sample
to 30 arcminutes in radius from the cluster center (the maximum spatial
coverage of the Hydra instrument on the WIYN 3.5m telescope).  For these
stars, we estimate $V$ magnitudes from 2MASS photometry, using an
empirical relationship $V=J+2.46(J-K)+0.40$, valid for the region of
NGC~6819.  
To derive this relationship, we used about 850 common stars between 2MASS
catalog and Phot03, all brighter than J=15. The standard deviation of the
fit is 0.12 mag. This transformation provides a rather crude estimate
of V magnitudes.  It is used only for 1.7\% of stars missing BV photometry
over the inner 10x10 arcmin area. In the outer parts of our field,
recently added to our survey, 
the fraction of stars without BV photometry can be as high as 100\%.
We discuss the development of our stellar sample for the WIYN RV
survey in \S~\ref{def_primary}.

\subsection{Astrometric Coordinates}

For all four photometric studies, we have also reduced all original
pixel data into sky coordinates using the UCAC2 \citep{zacharias04},
an accurate, dense, and relatively deep ($r_{lim}\approx 16$)
astrometric catalog. All datasets require quadratic and cubic terms in
the astrometric plate model. The standard error of astrometric
solutions ranges from 50 to 100 mas, with the higher end of the errors
attributed to the CFHT 3.6m telescope's CFH12K CCD mosaic data (K01).

We use the 2MASS Point Source Catalog as the primary catalog of stars in
the NGC~6819 field, to which we cross-correlate the Phot98, Phot03 and
UCAC2 catalogs of photometry and astrometry. In averaging positions, a
particular source is excluded only if it shows an offset exceeding 300
mas. The combination of these catalogs provides a comprehensive database
for NGC~6819 out to a $30\arcmin$ radius, effectively based on the UCAC2
coordinate system. These final coordinates are provided in Table
\ref{the-table}. The precision of mean positions is about 30 mas.

\subsection{WOCS Numbering System}
\label{s:wocsid}

Here we introduce the WOCS numbering system, based on $V$ magnitude
and radial distance from the cluster center. Separate one-dimensional
Gaussian fits in right ascention (RA) and declination (Dec) to the
cluster's apparent density profile (for stars with membership
probability $p>50\%$) provides the following new J2000 center:
$\alpha=19^{\rm h}41^{\rm m}17\fs5$
$\delta=+40\arcdeg11\arcmin47\arcsec$.  Around this center, annuli of
$30\arcsec$ width are drawn, and in each ring the stars are sorted
in increasing order of their $V$ magnitudes (i.e., the brightest stars
have the lowest numbers).  If a star is missing a measurement of its
$V$ magnitude, it is estimated from the 2MASS catalog using the
empirical relationship defined above.  The identification number (ID)
is the three digit star number followed by the three digit annulus
number, (e.g., the star 001003 is the brightest star in the third
annulus).\footnote{We note that the $V$ magnitude used to order the
  stars is not the final catalog $V$ magnitude, which was derived at a
  later stage of reductions. Thus the sequence may not be precisely
  consistent with the photometry.}

\section{Stellar Sample for the Radial-Velocity Survey}
\label{s:samp}

To improve observational efficiency, previous RV surveys of open
clusters have often pre-selected their target stars with preference to
proper-motion members.  Unfortunately, NGC 6819 lacks a complete
proper-motion database from which we can efficiently prioritize our
sample.  For a rich cluster in the Galactic plane, such as NGC~6819, the
observational requirements for a complete, unbiased RV survey are
particularly daunting. However, the capabilities of modern multi-object
spectrographs have grown to such a degree that we have been able to
measure RVs for a large sample of stars in the field of NGC~6819 in the
absence of measured proper motions.  Our full NGC 6819 RV database is
the combination of a WIYN and a CfA data set; we describe the respective
stellar samples below.

\begin{figure*}[!ht]
\plotone{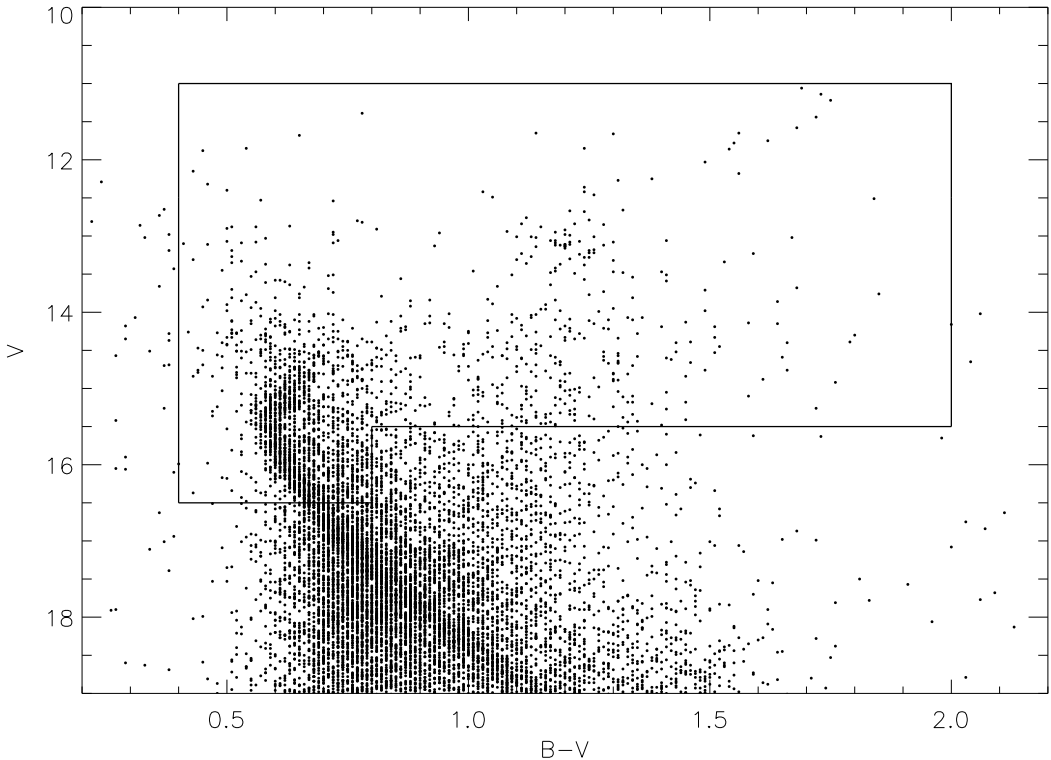}
\caption{
\label{cap:CMD-of-stars}
CMD displaying all stars in a 30-arcminute-radius field around NGC~6819.
The outlined region designates the photometric limits of our primary
WIYN stellar sample.
}
\end{figure*}

\subsection{WIYN}
\label{def_primary}

When we began our WIYN RV survey of NGC 6819 in 1998, we used the
photometry from RV98 and Phot98 to compile our ``primary sample''.  Thus
this sample of stars is limited in spatial extent to the 23 arcminute
field-of-view of the Phot98 observations.  Our RV observations are most
complete within the primary sample (see \S~\ref{ss:completeness}), as we
have only recently extended our survey to include additional stars
observed in subsequent photometric studies.  In the following section,
we describe the development of our stellar sample over the $\sim$10
years of our RV survey.

Our selection for the WIYN primary sample was influenced by the
instrument used for the study, the Hydra Multi-Object Spectrograph (MOS)
on the WIYN 3.5 m telescope on Kitt Peak \citep{barden94}.  The Hydra MOS has an
effective dynamic range of roughly four magnitudes within a given configuration of fibers, or a 
``pointing''; sources more than four magnitudes fainter than the brightest
are vulnerable to contamination by scattered light in the
spectrograph.  The faintest sources that Hydra MOS can observe
efficiently at high spectral resolution are $V \sim 16.5\
\mathrm{mag}$; this therefore sets our faint limit in magnitude.  At
colors bluer than $(\bv)\sim$0.4 (or $(B-V)_o \sim 0.2\ \mathrm{mag}$
using the $E(B-V) = 0.16\ \mathrm{mag}$ found by RV98), lower line
densities and increased line widths, typical of earlier-type stars,
make deriving reliable RV measurements increasingly difficult.
Therefore our blue limit in color is set by astrophysical constraints.

We therefore chose to define our WIYN primary sample to cover the
magnitude range of 11$\leq$$V$$\leq$15.5 within the color range of
0.4$\leq(\bv)\leq$2.0, with an additional magnitude range of
15.5$\leq$$V$$\leq$16.5 within the color range of 0.4$\leq(\bv)\leq$0.8.
Our photometric selection criteria are shown in the CMD of 
Figure~\ref{cap:CMD-of-stars}.  The reduced
color range for stars on the fainter end is designed to concentrate
attention on the main sequence at magnitudes below the cluster turnoff
and subgiant branch. As our primary sample was originally compiled from
the Phot98 photometric study, this sample covers the same 23 arcminute
square field-of-view.  The primary sample contains \numprimestars~stars,
covering the upper main sequence through the turnoff region and the
giant branch, including the red clump, as well as most potential blue
stragglers. This initial list of target stars has since been increased
by the addition of subsequent photometric studies and 2MASS sources
within our magnitude range (as described in \S~\ref{s:phot}).  However,
our observations are most complete within the primary sample, and, as
such, most of the results presented in the paper are derived from this
sample.

\subsection{CfA}

Mathieu and Latham began observations of NGC~6819 to measure RVs at the
Harvard-Smithsonian Center for Astrophysics facilities in 1988. A sample
of \numcfaall\ stars were selected from the proper-motion study of
\citet{sanders72} and the photometric survey of \citet{auner74}.  Given
the effective magnitude limit of the CfA Digital Speedometers, that
study was only able to reach the top of the main sequence. \numcfaprime\
of these \numcfaall\ stars are within the WOCS primary sample.  The
remainder are generally brighter than $V = 11.0\ \mathrm{mag}$, bluer
than $(B-V)=0.4$, or have incomplete photometry.

\section{Radial-Velocity Observations, Data Reduction and Precision }
\label{odp}

\subsection{WIYN}

The Hydra MOS is a fiber-fed spectrograph with a one-degree field of
view, currently capable of taking $\sim$80 simultaneous spectra, with an effective
dynamic range of roughly four magnitudes within a given pointing.  In total, we
have observed 90 pointings on NGC 6819 over 35 separate observing runs
on the WIYN 3.5m.

Developing a strategy for prioritizing the stars in our sample for
placement of the $\sim$80 fibers during each pointing is critical to
optimizing limited observing time.  In order to satisfy the
four-magnitude dynamic range, we first define a faint sample, covering
the magnitude range of 12.5$\leq$$V$$\leq$16.5, to be observed in good
weather conditions. We also developed a bright sample, covering the
magnitude range of 11$\leq$$V$$\leq$15, to be observed in the case
when light cloud cover would likely prevent us from deriving reliable
RVs for fainter stars.  Our original strategy for observations gave
highest priority to stars within $200\arcsec$ ($\sim$2 pc) from the
cluster center; within that radius, targets were prioritized by
brightness. Outside $200\arcsec$, where membership probability
decreases significantly with radius, sources were prioritized by
radius from the cluster center.

As our survey matured, we adopted a more sophisticated strategy for
prioritizing the stars in our observing lists (both faint and bright).
Monte Carlo simulations show that we require at least three observations
over the course of a year to ensure 95\% confidence that a star is
either constant or variable in velocity \citep{mathieu83}.  Given three
observations with consistent velocity measurements over a timespan 
of at least a year and typically longer, we classify a given
star as single (strictly, non-velocity variable) and finished, and move
it to the lowest priority. If a given star has three RV measurements
with a standard deviation $>$1.6 km s$^{-1}$ (four times our precision;
see \S~\ref{vvar}), we classify the star as velocity variable and give
it the highest priority for observation on a schedule appropriate to its
timescale of variability.  This prioritization allows us to most efficiently
derive orbital solutions for our detected binaries.
We have made a strong effort to observe all
stars in our primary sample with $V\leq$15 at least three times, and
have therefore prioritized these stars in our observations.  These stars
span the giant branch through the upper-main sequence and contain most
potential blue stragglers.  There are \numprimebright\ stars within our
primary sample with $V\leq$15.

We place our shortest period binaries at the highest priority for
observations each night, followed by longer period binaries to obtain
1-2 observations per run.  Below the confirmed binaries we place, in
the following order, ``candidate binaries'' (once-observed stars with
a radial-velocity measurement outside the cluster radial-velocity
distribution or stars with a few measurements that span only 1.5-2 km
s$^{-1}$), once observed and then twice observed non-velocity-variable
likely members, twice observed non-velocity-variable likely
non-members, unobserved stars, and finally, ``finished'' stars.
Within each group, we prioritize by distance from the cluster center,
giving those stars nearest to the center the highest priority.

Our observing procedure and data reduction process for WOCS RV
observations are described in detail in \citet{geller08}.  Briefly, we
use the echelle grating, providing a spectral resolution of roughly
$15\ \mathrm{km\ s^{-1}}$. The majority of our spectra are centered at $513.0\
\mathrm{nm}$ with a range of $25\ \mathrm{nm}$, covering numerous
narrow absorption lines including the Mg I b triplet.  During data
reduction, the images are bias- and sky-subtracted, and the extracted
spectra are flat fielded, throughput corrected, and dispersion
corrected.  The WIYN spectra have signal-to-noise ratios ranging from 
$\sim$18 per resolution element for $V$=16.5 stars to 
$\sim$120 per resolution element for $V$=12.5 stars in a two-hour exposure.
The RVs are derived from a one-dimensional cross-correlation
with an observed solar template spectrum, corrected to be at rest 
\citep[e.g.,][]{tonry79}.  
We performed a detailed study of the effect of using the solar template across our
$(\bv)$ color range in \citet{geller08}, finding no
noticable systematic offset (given our precision of 0.4 km s$^{-1}$).
These RVs are then converted
to heliocentric RVs and are corrected for the unique fiber offsets of
the Hydra MOS.

We have analyzed the precision of our WIYN NGC 6819 RV measurements in
the same method as in \citet{geller08}, following the process
described in \citet{kirillova63}.  A $\chi^2$ function fit to the
distribution of standard deviations of our NGC 6819 WIYN RV
measurements yields a precision for our WIYN data of $\precision\
\mathrm{km\ s^{-1}}$ for a single observation, the same value found in
\citet{geller08}.

\subsection{CfA}

The CfA RV measurements were obtained with two nearly identical
instruments on the Multiple Mirror Telescope\footnote{\MMTfoot} and the
1.5-m Tillinghast Reflector at the Whipple Observatory atop Mt. Hopkins,
Arizona \citep{latham92}. Echelle spectrographs were used with
intensified photon-counting Reticon detectors to record about $4.5\
\mathrm{nm}$ of spectrum in a single order near $518.7\ \mathrm{nm}$,
with a resolution of $8.3\ \mathrm{km\ s^{-1}}$ and signal-to-noise
ratios ranging from 8 to 15 per resolution element. Information on the
CfA data reduction process can be found in \citet{stefanik99}.

A $\chi^2$ analysis of the CfA precision yields a value of $\sim0.7\
\mathrm{km\ s^{-1}}$.  This value agrees well with that derived by 
\citet{mathieu86b} for CfA RVs from stars in M67.

\section{The Combined WIYN and CfA Radial-Velocity Data Set}

The majority of our observations, \numwiynmeasurements\ measurements of
\numwiynobserved\ stars, were taken with the WIYN Hydra MOS, starting in 
June 1998 and still ongoing.  
The WIYN measurements have a typical frequency of $\sim$four
epochs per year.  Additionally, we have \numcfameasurements\ CfA measurements
of \numcfaobserved\ stars. The bulk of the CfA measurements were taken
from May 1988 through October 1992, though some were taken through
1995.  The CfA measurements have a typical frequency of $\sim$five epochs
per year.

\begin{figure*}[!ht]
\plotone{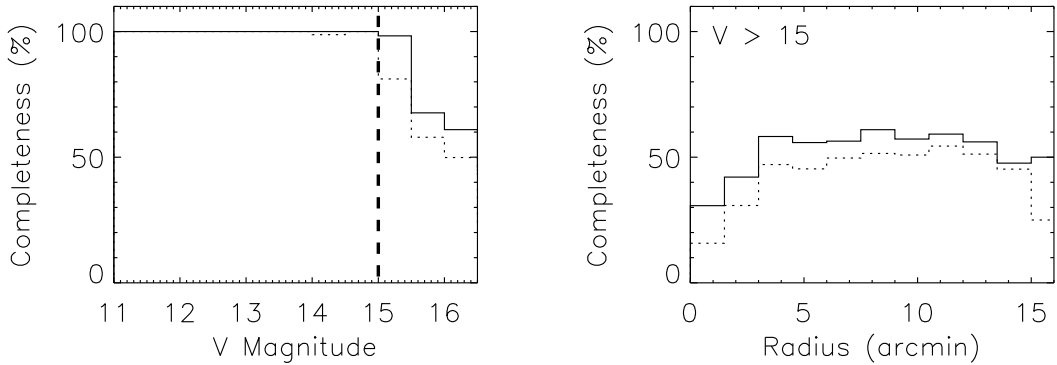}
\caption{Completeness histograms of observed stars in our NGC 6819
primary sample.  For both plots, we use the solid and dashed lines to
indicate the completeness in stars with $\geq$1 and $\geq$3
radial-velocity measurements, respectively.  On the left, we show the
completeness as a function of $V$ magnitude.  We are essentially
complete for all stars with $V \leq$ 15, as indicated by the dashed
line.  On the right, we show the completeness for stars with $V >$ 15,
essentially plotting our completeness amongst the main sequence stars in
our sample, as a function of radius from the cluster center.  }
\label{compfig}
\end{figure*}

Prior to combining the WIYN and CfA data sets, we first searched for a
potential zero-point offset by comparing stars with $\geq 3$
measurements in each sample and with a standard deviation of $\leq 1.0\
\mathrm{km\ s^{-1}}$.  There are 15 such stars common to both the WIYN
and CfA samples. One of these stars has a difference in average
velocities of $23\ \mathrm{km\ s^{-1}}$ and was removed from the
comparison. For the remaining stars, the mean offset between the WIYN
and the CfA average velocities is $0.07\ \mathrm{km\ s^{-1}}$, less than
the standard deviation of the mean of the difference ($0.11\
\mathrm{km\ s^{-1}}$).  As stated above, the precision on our WIYN
measurements is \precision\ km s$^{-1}$, and the precision on our CfA
measurements is approximately twice that, at 0.7 km s$^{-1}$.  Thus, we
conclude that there is no significant zero-point offset between the two
data sets at the level of our precision, and we therefore combine the
WIYN and CfA data without modification.  When using the measurements in
our analyses, such as RV averages, they are weighted by the inverse of
their respective precisions (see Equation~\ref{weighteq2}).  We note that the CfA data significantly
increase our time baseline with which to detect and find orbital
solutions for long-period binaries.

\subsection{Completeness Within the Primary Sample}
\label{ss:completeness}

Of the \numprimestars\ stars in our primary sample (see
\S~\ref{def_primary}) we have at least one RV measurement of \numprimeobserved\ 
targets (\percprimeobserved). Over half of the
stars in this primary sample have sufficient measurements for their RVs
to be considered final 
(\numprimefinal\ of \numprimestars, \percprimefinal). 
This means that for each of these stars we have at
least three velocity measurements that are consistent, or if they are
variable, that we have found binary orbital solutions (discussed in \S
\ref{orb_soln}).  Of those stars not finalized, 130
stars have only one or two observations and another
\numprimebinnosolns\ stars are variable but do not yet have definitive
orbital solutions. Because of our emphasis on the brighter (i.e. more
evolved) stars (\S \ref{odp}), we have final measurements for 386
of the 436
stars in the primary WIYN sample with $V \leq 15.0\ \mathrm{mag}$, 
for a completeness of 89\%.
Forty eight
of the remaining 50
stars are velocity variables (including some rapidly rotating stars) without 
orbital solutions yet.

In Figure~\ref{compfig} we plot the completeness in our primary sample
as a function of both $V$ magnitude (left) and projected radius (right).
We plot the completeness in stars observed $\geq$3 times with the dashed
line, and stars observed $\geq$1 time with the solid line.  A targeted effort has been made to ensure that our observations for this sample are nearly complete down to $V$=15; there
are only two
stars that have less than three observations
in this bright sample.  Both are rotating too rapidly to derive reliable RVs 
with our current observating setup.  
Towards fainter magnitudes, the completeness
drops to $\sim$60\% at $V$=16.5.  These fainter stars require clear, dark skies in order to
derive reliable RV measurements, and there is a very large increase in the number of
stars in our primary sample as we begin to include the main sequence
(Figure~\ref{cap:CMD-of-stars}). 

Further, crowding limits for the Hydra
MOS fibers and the high surface density of main sequence stars near the
cluster center make it more challenging to obtain the same completeness
in this region. This can be seen in Figure~\ref{compfig} showing completeness 
as a function of radius for (largely) main-sequence stars.

\section{Results}
\label{s:results}

\begin{deluxetable*}{cccccccccccccc}
\tabletypesize{\footnotesize} 
\tablewidth{0pc} 
\tablecolumns{14}
\tablecaption{
The first ten lines of the NGC 6819 WOCS radial-velocity database. 
Coordinates are in J2000. The radial velocity given is the mean of the
measurements.  For binaries with solutions, we include the
center-of-mass velocity $\gamma$ and any comments about the nature of
the system. The standard error of the mean is also given. If we only
have a single measurement, the WIYN or CfA measurement precision is
given instead, as appropriate.  For binaries with orbital solutions, we
provide the error on $\gamma$ from the orbital fit in this column
instead.
\label{the-table}}
\tablehead{
\colhead{WOCS} &
\colhead{} &
\colhead{} &
\colhead{} &
\colhead{} &
\colhead{} &
\colhead{} &
\colhead{} &
\colhead{Std. } &
\colhead{} &
\colhead{Mem. } &
\colhead{} &
\colhead{} &
\colhead{} 
\\[-2pt]
\colhead{ID\tablenotemark{a} } &
\colhead{$\alpha$ } &
\colhead{$\delta$ } &
\colhead{V} &
\colhead{B-V} &
\colhead{$N_{WIYN}$ } &
\colhead{$N_{CfA}$ } &
\colhead{$\overline{RV}$ } &
\colhead{Err. } &
\colhead{e / i } &
\colhead{Prob. } &
\colhead{Class\tablenotemark{b} }  &
\colhead{$\gamma$}  &
\colhead{Comment} 
}
\startdata
001001 & 19 41 18.71 & 40 11 42.9 & 12.68 & 1.24 & 3 & 20 &  6.04      &  3.40     &  23.88 & \nodata & BU &  \nodata  &  \nodata  \\ 
002001 & 19 41 18.93 & 40 11 41.0 & 13.38 & 1.24 & 3 & 0 &  0.94      &  0.04     &  0.13     & 88 & SM &  \nodata  &  \nodata  \\ 
003001 & 19 41 18.76 & 40 11 54.8 & 13.64 & 1.17 & 4 & 2 &  1.42      &  0.30     &  1.29     & 93 & SM &  \nodata  &  \nodata  \\ 
004001 & 19 41 15.75 & 40 11 36.0 & 13.98 & 1.19 & 3 & 5 &  1.49      &  0.18     &  0.77     & 93 & SM &  \nodata  &  \nodata  \\ 
005001 & 19 41 16.21 & 40 11 26.4 & 15.02 & 0.64 & 1 & 0 &  4.90      &  0.4 &  \nodata & \nodata & U &  \nodata  &  \nodata  \\ 
006001 & 19 41 18.25 & 40 11 34.8 & 15.20 & 0.61 & 2 & 0 &  -0.93     &  1.87     &  \nodata & \nodata & U &  \nodata  &  \nodata  \\ 
007001 & 19 41 17.27 & 40 11 27.3 & 15.20 & 0.65 & 3 & 0 &  2.24      &  0.28     &  1.00     & 95 & SM &  \nodata  &  \nodata  \\ 
008001 & 19 41 18.63 & 40 11 35.2 & 15.27 & 0.61 & 3 & 0 &  2.16      &  2.27     &  8.04 & \nodata & BLM &  \nodata  &  \nodata  \\ 
009001 & 19 41 18.01 & 40 11 19.6 & 15.48 & 0.61 & 1 & 0 &  5.66      &  0.4 &  \nodata & \nodata & U &  \nodata  &  \nodata  \\ 
011001 & 19 41 16.51 & 40 11 53.4 & 15.73 & 0.60 & 1 & 0 &  3.73      &  0.4 &  \nodata & \nodata & U &  \nodata  &  \nodata  \\ 

\enddata
\tablenotetext{a}{See \S~\ref{s:wocsid} for an explanation of the WOCS ID number.}
\tablenotetext{b}{See \S~\ref{class} for an explanation of the class codes.}
\end{deluxetable*}

Our full NGC 6819 database is available with the electronic version of
this paper; here we show a sample of our results in Table
\ref{the-table}.  For each star, we list the WOCS identification
number, right ascension ($\alpha$), declination ($\delta$), the
averaged $BV$ photometry (see \S \ref{s:phot}), number of RV
measurements, the mean and standard error of the RV measurements, the
e/i value (see \S~\ref{vvar}), the calculated RV membership
probability (see \S~\ref{s:prob}), and the classification of the
object (see \S~\ref{class}). For velocity-variable stars with orbital
solutions, we present the center-of-mass ($\gamma$) RV and its
standard error, and add the comment SB1 or SB2 for single- and
double-lined binaries, respectively.

\subsection{Radial-Velocity Membership Probabilities}

\label{s:prob}

\begin{figure}[!ht] 
\plotone{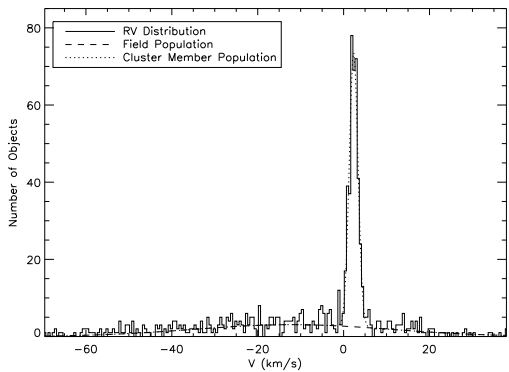}
\caption{ 
\label{cap:vel-distrib} 
RV distribution of single stars with $\geq$3 RV measurements. 
The cluster population is clearly distinguishable from the field
distribution as the tightly peaked distribution centered on
$\clustrvmean\ \mathrm{km\ s^{-1}}$. }
\end{figure}

NGC~6819 lies close to the plane of the Galaxy ($\ell=74\fdg0$
$b=+8\fdg5$), and the cluster RV distribution is embedded within the
field velocity distribution. Even so, in a histogram of the RVs of
single stars, the cluster population is readily distinguishable from
the bulk of the field stars (Figure~\ref{cap:vel-distrib}). The
cluster can be seen as the tightly peaked velocity distribution
($\sigma \sim \clustrvdev\ \mathrm{km\ s^{-1}}$) centered around a
mean velocity of $\clustrvmean\ \mathrm{km\ s^{-1}}$. In order to
calculate RV membership probabilities for each star, we simultaneously
fit one-dimensional Gaussian functions $F_{c}(v)$ and $F_{f}(v)$ to
the cluster and field RV distributions respectively. We then compute
membership probability $p(v)$ with the usual formula:
\begin{equation} \label{memeq}
p(v) = \frac{F_{c}(v)}{F_{f}(v)+F_{c}(v)}  
\end{equation}
\citep{vasilevskis58} (see Table~\ref{t:fit} for fit parameters).  
We use only single stars in computing the Gaussian fits.  The RV
distribution and the Gaussian fits are shown in
Figure~\ref{cap:vel-distrib}.

\begin{figure}[!ht]
\plotone{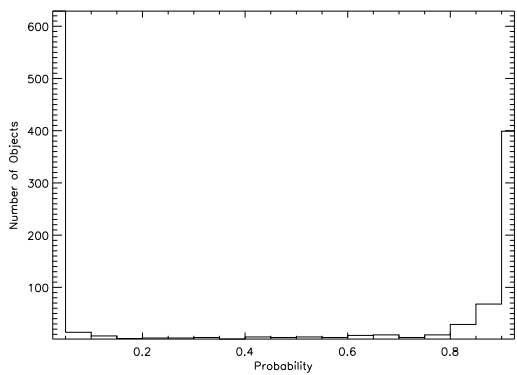}
\caption{ 
\label{f:memb_prob} 
RV membership probabilities for single stars with $\geq$3 RV measurements in the field of view of
NGC~6819. The probabilities display a clean separation between cluster
and field stars.  We classify as cluster members those stars with a RV
membership probability $p \geq 50\%$.  }
\end{figure}

For single stars, we use the mean RVs in Table~\ref{the-table} to
compute membership probabilities.  For binary stars with orbit solutions, we compute membership probabilities from the $\gamma$ velocity. For velocity-variable stars without
orbit solutions, the $\gamma$-velocities are not known, and therefore we
cannot calculate an RV membership.  For these stars, we provide a
preliminary membership classification, described in \S~\ref{class}.

The probability distribution in Figure~\ref{f:memb_prob} shows a very
clean separation of cluster members and field stars. In the following
analysis, we use a probability cutoff of $p \geq 50\%$ for classification
as a member. This criterion for membership gives \numsingmembers\ single
cluster members.  Using only these single
members, we find a mean cluster velocity of \fancyclustcent\ $\pm$
\fancyclusterr\ $\mathrm{km\ s^{-1}}$.  From the area under the fit to
the cluster and field distributions, we expect \expsingmem\ single
cluster members as well as \expcontam\ field stars having velocities
that result in $p \geq 50\%$.  Thus we estimate a field contamination of
\perccontam . Though this estimate is derived from single stars, the
percent contamination should be valid for the cluster as a whole.

\begin{deluxetable}{c r@{\hspace{0.5em}}c@{\hspace{0.5em}}l r@{\hspace{0.5em}}c@{\hspace{0.5em}}l}
\tablewidth{0pc}
\tablecolumns{3}
\tablecaption{Gaussian Fit Parameters For Cluster and Field RV Distributions
\label{t:fit}}
\tablehead{\colhead{} & \multicolumn{3}{c}{Cluster} & \multicolumn{3}{c}{Field}}
\startdata
Ampl. (Number) & 58.3 &$\pm$& 1.0  & 3.1 &$\pm$& 0.3 \\
$\overline{RV}$ (km\ s$^{-1})$ & 2.338 &$\pm$& 0.019 & -12.0 &$\pm$& 1.8 \\
$\sigma$ (km\ s$^{-1})$ & 1.009 &$\pm$& 0.019 & 23 &$\pm$& 3
\enddata
\end{deluxetable}

\subsection{Velocity-Variable Stars}
\label{vvar}
\label{orb_soln}

Velocity-variable stars are distinguishable by the larger standard
deviations of their RV measurements.  Here, we assume that such velocity
variability is the result of a binary companion (or perhaps multiple
companions).  Specifically, we consider a star to be a velocity-variable
if the ratio of the standard deviation of its RV measurements to our
measurement precision is greater than four \citep{geller08}.  We refer
to this ratio as e/i, where ``e'' is the standard deviation of the RV
measurements for the star, and ``i'' is our measurement precision.  As
stated in \S~\ref{odp} we find a precision for the WIYN data of
$\precision\ \mathrm{km\ s^{-1}}$ while the CfA data have a precision of
0.7 km s$^{-1}$.  For stars with multiple RV measurements from
both observatories, our combined e/i value is weighted by the expected
precision of each measurement.  From \citet{bevington69}, the variance for a data
set with multiple precision values is defined as
\begin{equation}
e^2 = \frac{N}{N-1}\frac{\displaystyle\sum_i^N (RV_i - \bar{RV})^2 / \sigma_i^2}{\displaystyle\sum_i^N 1/\sigma_i^2} .
\end{equation}
The square of the expected precision for this data set is defined as
\begin{equation}
i^2 = \frac{1}{N}\displaystyle\sum_i^N \sigma_i^2 .
\end{equation}
Thus the e/i value for stars with multiple RV measurements from both 
observatories is given by 
\begin{equation} \label{weighteq1}
\left(\frac{e}{i}\right)^2 = 
\frac{N^2}{N-1} \frac{\displaystyle\sum_i^N (RV_i - \bar{RV})^2 / \sigma_i^2}
{\displaystyle\sum_i^N \sigma_i^2 \displaystyle\sum_i^N 1/\sigma_i^2} 
\end{equation}
where the $\bar{RV}$ is the mean RV weighted by the respective precision values, 
and is defined as,
\begin{equation} \label{weighteq2}
\bar{RV} = \frac{\displaystyle\sum_i^N (RV_i/\sigma_i^2)}{\displaystyle\sum_i^N (1/\sigma_i^2)}.
\end{equation}
Again, $\sigma_i$ is 0.4 km s$^{-1}$ for WIYN measurements and 0.7 km s$^{-1}$ 
for CfA measurements. 


In this manner, we can calculate reliable e/i values for narrow-lined
stars.  Stars with e/i$<$4 are labelled as single; however, certainly
some fraction of these stars are long-period and/or low-amplitude
binaries. For double-lined binaries and rapidly-rotating stars,
however, the precision on our measurements is less well defined. We do
not derive an e/i value for such stars. In the case of double-lined
spectra, we take their multiplicity as given and label them as 
velocity variables directly (providing the comment of SB2 in Table~\ref{the-table}).

To date, we have identified \numbinall\ velocity variables. We have
derived orbital solutions for \numbinsolns\ of these. \numbinCOMmem\ of
the resulting gamma velocities give $p \geq 50\%$ and are thus likely to
be cluster members.  For the binaries with orbital solutions, we quote
the $\gamma$-velocities in Table~\ref{the-table}.  In following papers,
we will provide the full orbital solutions including all derived
parameters for each binary, as well as detailed analyses of the
distributions of orbital parameters and the binary frequency of the
cluster.

\subsection{Membership and Variability Classification}
\label{class}

In addition to our RV membership probabilities and e/i measurements, we
also provide a qualitative classification for each narrow-lined star
observed $\geq$3 times as a guide to its membership and variability.
Again, we consider a star to be single if its e/i$<$4.  For these stars
we classify those with $p \geq 50\%$ as single members (SM), and those
with $p < 50\%$ as single non-members (SN).  If a star has e/i$\geq$4 and
enough measurements from which we are able to derive an orbital
solution, we use the $\gamma$-velocity to compute a secure membership.
For these binaries, we classify those with $p \geq 50\%$ as binary
members (BM) and those with $p < 50\%$ as binary non-members (BN).  For
velocity variables without orbital solutions, we split our
classifications into three categories.  If the mean RV results in $p
\geq  50\%$, we classify the system as a binary likely member (BLM).  If
the mean RV results in $p < 50\%$ but the range of measured velocities
includes the cluster mean velocity, we classify the system as a binary
with unknown membership (BU).  Finally, if the RV measurements for a
given star all lie either at a lower or higher RV than the cluster
distribution, we classify the system as a binary likely non-member
(BLN), since it is unlikely that any orbital solution could place the
star within the cluster distribution.  We classify 
stars with $<$3 RV measurements as well as some rapid rotators as unknown (U), 
as these stars do not
meet our minimum criterion for deriving RV memberships or e/i
measurements.  In the following analysis, we include the SM, BM and BLM
stars as cluster members.  We list the number of stars within each class
in Table~\ref{t:cat}. The total number of cluster members in our sample
is
\nummembers .

\begin{deluxetable}{cc}
\tablewidth{0pc}
\tablecolumns{2}
\tablecaption{Number of Stars Within Each Classification
\label{t:cat}}
\tablehead{ \colhead{Classification} & \colhead{N Stars}}
\startdata
SM & \numcatSM \\
SN & \numcatSN \\
BM & \numcatBM \\
BN & \numcatBN \\
BLM & \numcatBLM \\
BLN & \numcatBLN \\
BU & \numcatBU \\
U & \numcatU \\
\enddata
\end{deluxetable}

\section{Discussion}
\label{disc}

\subsection{Color-Magnitude Diagram}

\begin{figure}[!ht]
\plotone{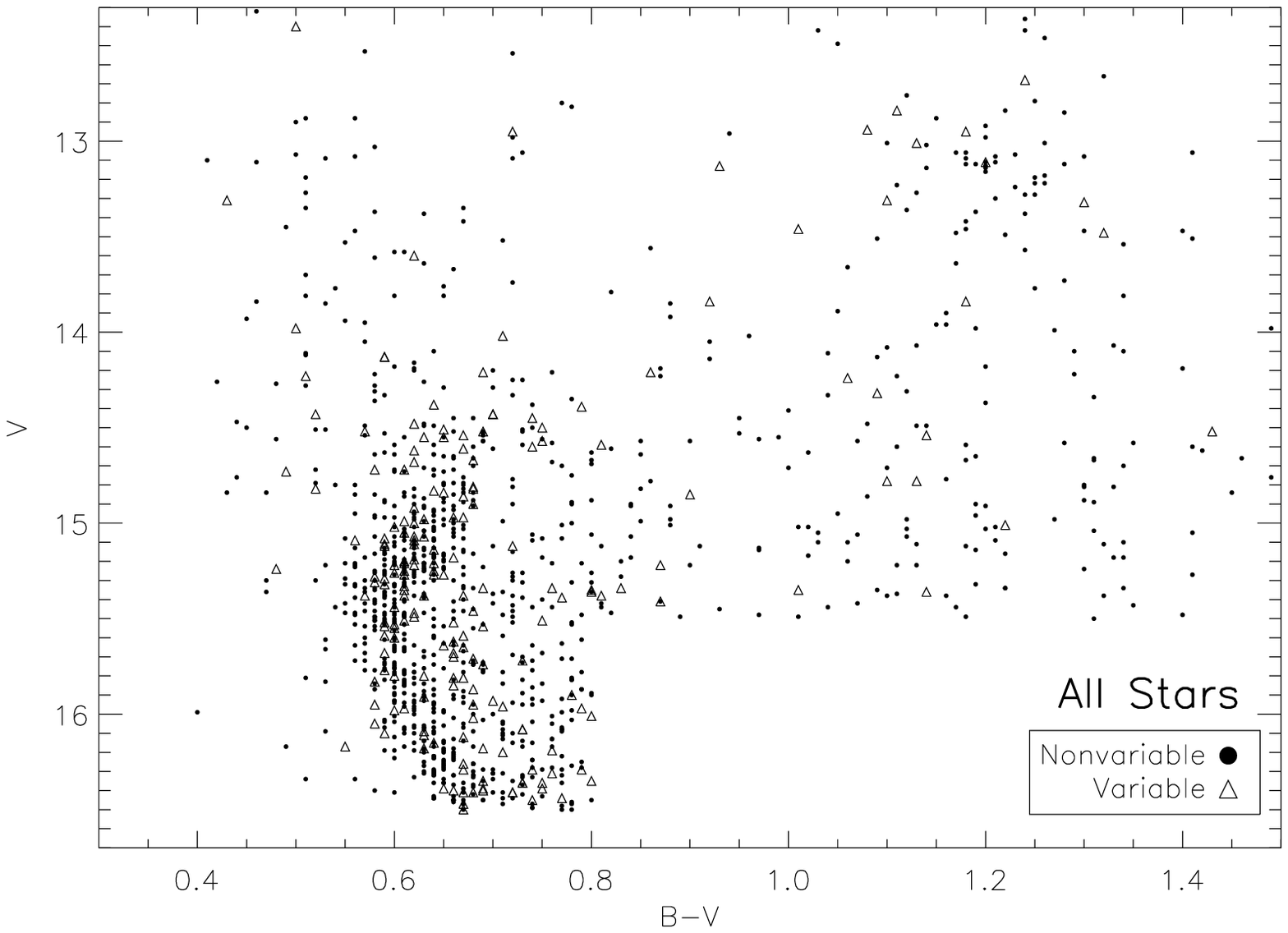}
\plotone{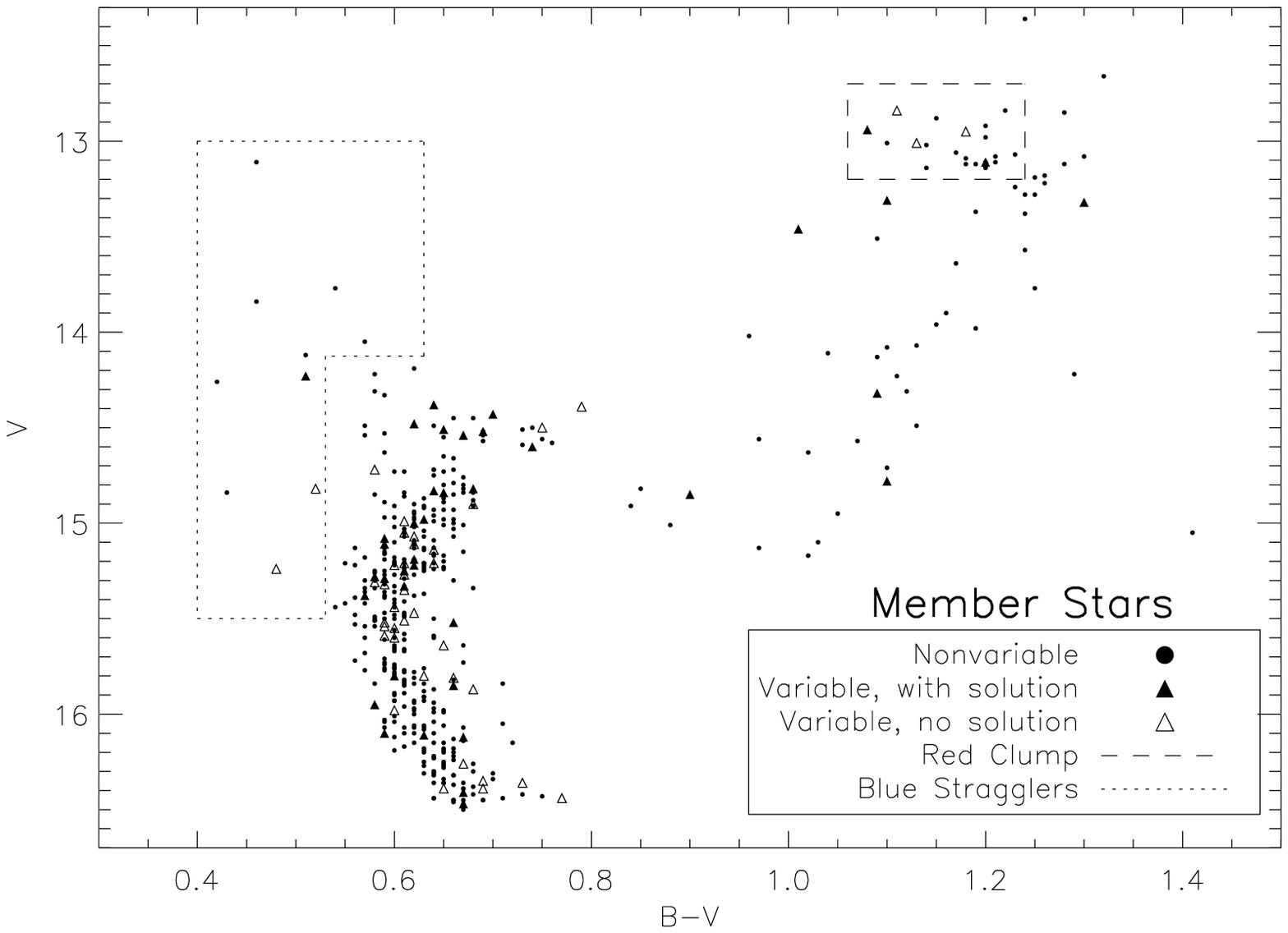}
\caption{
\label{all-mem-cmds}
Color-magnitude diagram of all stars with WOCS RV measurements (top) and
RV-selected narrow-lined member stars (bottom). The triangles in the upper diagram
represent velocity variables. The filled triangles in the lower (cluster
member) diagram are binaries with orbital solutions (BM). Open triangles
are velocity variables currently without solutions but with mean
velocities indicating membership (BLM). The color and magnitude criteria
used for identifying the red clump and blue straggler populations are also
indicated with the dotted and dashed lines, respectively. }
\end{figure}

The ability of our RV survey to distinguish between members of NGC 6819
and the field is evident in a comparison of the two CMDs shown in
Figure~\ref{all-mem-cmds}. The upper CMD includes all stars for which we
have RV measurements, while the lower CMD includes only RV-selected
cluster members (i.e., SM, BM and BLM stars).  In the latter, the
classic cluster sequence is revealed from the upper main sequence
through the red clump.  We plot the velocity variables with triangles
and the single stars with circles.  Note the rich giant, subgiant and
blue straggler populations as well as the large number of detected
binaries.

\begin{figure}[!ht]
\plotone{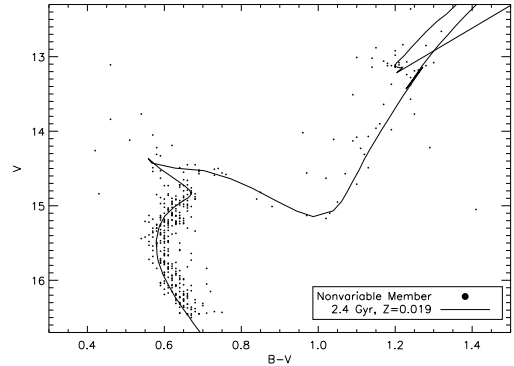}
\caption{
\label{isos}
CMD showing only single narrow-lined
cluster members.  The cluster displays well-populated subgiant and giant
branches and a population of single blue stragglers brighter and bluer than the cluster turnoff. The main sequence shows a marked bend to the red at the turnoff
and a gap between the main-sequence turnoff and the subgiant branch. Also shown is an isochrone from the core convective overshoot models of \citet{marigo08} for reference. The turnoff morphology is fit well by the isochrone, and is characteristic of core convective overshoot. }
\end{figure}

In Figure~\ref{isos} we again present the CMD with only kinematically-selected, single
cluster members.  In this figure we also plot 
a \citet{marigo08} isochrone model, 
which includes core convective overshoot. 
The displayed
isochrone was created using
cluster age of \clusterage\ Gyr and solar metallicity, which are consistent with the range of cluster parameters determined by different groups for NGC 6819. The isochrone was placed by eye, using $(m-M) = $
12.3 mag
and $E(B-V) = $
0.1 mag.
With this isochrone the cluster turnoff is found to be at 
$\sim 1.53\ M_{\sun}$.

There is mounting evidence that the Schwarzschild criterion
($\nabla_{ad} \geq \nabla_{rad}$) underestimates the size of the
convective core of intermediate mass stars, most likely due to effects
of turbulence and rotation at the convective-radiative boundary 
\citep[e.g.,][]{shaviv73,zahn02}. This
provides extra fuel to the hydrogen-burning core and increases the star's main 
sequence lifetime, as well as the
eventual size of the hydrogen-depleted region when core fusion
ceases. As the star leaves the main sequence, the core begins to
contract, the temperature in the outer layers increases and eventually
hydrogen fusion ignites in a shell around the depleted core.  Because
the depleted region is larger in the convective overshoot model, the
contraction and ignition phases occur on a different timescale than they
would in a ``classical'' core, calculated using the Schwarzschild
criterion.  It is this more extended contraction time and more powerful ignition that results in the ``blue hook'' morphology of the main sequence turnoff.

As evident from the displayed isochrone, the observational signatures of convective overshooting include an enhanced bend to the red of the main sequence turnoff, a blue hook structure to the very top of the main sequence resulting from a rapid evolution of turnoff stars to the blue, and a gap between the top of the main sequence and the subgiant branch. Such a gap is clearly evident near the top of the NGC 6819 main sequence at roughly $V = 14.6\ \mathrm{mag}$. We note that to securely reveal such a gap it is imperative that, in addition to removing non-members, binaries be identified and removed, since a binary population can mimic such a gap (e.g., Figure 5). It remains possible that the population of stars on the brighter side of the gap includes wider equal-mass binaries not identifiable by our spectroscopic techniques.

Whether stars currently populate the blue hook itself is unclear because of confusion with the blue straggler population. Certainly there are stars in the NGC 6819 CMD consistent with populating the blue hook structure in the models. A study of the population density along the isochrone is merited to compare the expected and actual numbers. More detailed study of these stars may also allow identification of those that are blue stragglers, for example through differing rotation distributions. 

\subsection{Stars of Note}

\subsubsection{Blue Stragglers}

We mark the location of candidate blue stragglers within the dotted
lines in the bottom panel of Figure 5. 
This selection region is conservative, and does not include stars that 
could be currently populating the blue hook (either single stars or binaries),
based on the isochrone fit of Figure 6 (see also Section 7.1). Likely there are more blue stragglers
within the "notch" of the region.

Within this region we identify 12 candidate blue stragglers that are
bluer and brighter than the cluster turnoff. We include here four stars
that are to the blue of the turnoff, but somewhat fainter, as
interesting candidates for future study.

It is generally accepted that blue stragglers are, or were, members of
multiple systems whose evolution has been influenced by either stellar
evolution or through dynamical processes resulting in mass transfer,
mergers or even stellar collisions \citep[e.g.,][]{bailin95}.  We note
that only four of our potential blue stragglers currently display
velocity variability (including the rapid rotator 014012), as opposed to the
large frequency of binaries ($\sim$75\%) in the blue straggler 
population of the old (7 Gyr) open cluster NGC 188 \citep{geller08}.  
In Table~\ref{bstab} we list our data for these 
12 NGC~6819 blue stragglers, in the same format as in Table~\ref{the-table}.

\begin{deluxetable*}{cccccccccccccc}
\tabletypesize{\scriptsize}
\tablewidth{0pc}
\tablecolumns{14}
\tablecaption{NGC 6819 Candidate Blue Stragglers \label{bstab}}
\tablehead{\colhead{WOCS} & \colhead{} & \colhead{} & \colhead{} & \colhead{} & \colhead{} & \colhead{} & \colhead{} & \colhead{Std. } & \colhead{} & \colhead{Mem. } & \colhead{} & \colhead{} & \colhead{} \\[-2pt]
\colhead{ID} & \colhead{$\alpha$ } & \colhead{$\delta$ } & \colhead{V} & \colhead{B-V} & \colhead{$N_{WIYN}$ } & \colhead{$N_{CfA}$ } & \colhead{$\overline{RV}$ } & \colhead{Err. } & \colhead{e / i } & \colhead{Prob. } & \colhead{Class}  & \colhead{$\gamma$}  & \colhead{Comment}
}
\startdata

009003 & 19 41 17.01 & 40 10 34.8 & 13.77 & 0.54 & 5 & 1 &  2.60      &  0.19     &  0.93     & 95 & SM &  \nodata  &  \nodata  \\ 
014003 & 19 41 16.23 & 40 10 27.7 & 14.84 & 0.43 & 4 & 0 &  3.78      &  0.34     &  1.46     & 88 & SM &  \nodata  &  \nodata  \\ 
009005 & 19 41 28.22 & 40 12 54.5 & 14.12 & 0.51 & 5 & 1 &  0.73      &  0.48     &  2.31     & 85 & SM &  \nodata  &  \nodata  \\ 
010005 & 19 41 06.63 & 40 10 30.4 & 14.26 & 0.42 & 4 & 2 &  3.38      &  0.34     &  1.46     & 92 & SM &  \nodata  &  \nodata  \\ 
005006 & 19 41 28.99 & 40 13 15.5 & 13.84 & 0.46 & 4 & 1 &  2.18      &  0.47     &  2.00     & 95 & SM &  \nodata  &  \nodata  \\ 
016009 & 19 41 03.51 & 40 08 42.6 & 14.82 & 0.52 & 29 & 0 &  2.86      &  0.47     &  6.17 & \nodata & BLM &  \nodata  &  \nodata  \\ 
023011 & 19 40 54.75 & 40 08 35.5 & 15.44 & 0.54 & 6 & 0 &  2.45      &  0.12     &  0.66     & 95 & SM &  \nodata  &  \nodata  \\ 
014012 & 19 41 17.01 & 40 06 04.0 & 14.84 & 0.47 & 10 & 0 &  1.52      &  1.87     &  \nodata & \nodata & BLM &  \nodata  &  Rapid rotator  \\ 
007013 & 19 40 44.58 & 40 12 33.7 & 13.11 & 0.46 & 6 & 12 &  0.95      &  0.15     &  1.03     & 88 & SM &  \nodata  &  \nodata  \\ 
010016 & 19 40 56.00 & 40 18 39.1 & 14.23 & 0.51 & 23 & 0 &  2.63      &  0.04     &  9.08     & 95 & BM & 2.408 & SB1 \\ 
003022 & 19 40 48.32 & 40 02 28.9 & 13.10 & 0.41 & 8 & 1 &  2.35      &  0.28     &  \nodata & 95 & SM &  \nodata  &  Rapid rotator  \\ 
032023 & 19 40 54.90 & 40 01 07.8 & 15.24 & 0.48 & 20 & 0 &  2.47      &  0.43     &  4.70 & \nodata & BLM &  \nodata  &  \nodata  \\

\enddata
\end{deluxetable*}

\begin{deluxetable*}{cccccccccccccc}
\tabletypesize{\scriptsize}
\tablewidth{0pc}
\tablecolumns{14}
\tablecaption{NGC 6819 Candidate Red Clump Stars \label{rctab}}
\tablehead{\colhead{WOCS} & \colhead{} & \colhead{} & \colhead{} & \colhead{} & \colhead{} & \colhead{} & \colhead{} & \colhead{Std. } & \colhead{} & \colhead{Mem. } & \colhead{} & \colhead{} & \colhead{} \\[-2pt]
\colhead{ID} & \colhead{$\alpha$ } & \colhead{$\delta$ } & \colhead{V} & \colhead{B-V} & \colhead{$N_{WIYN}$ } & \colhead{$N_{CfA}$ } & \colhead{$\overline{RV}$ } & \colhead{Err. } & \colhead{e / i } & \colhead{Prob. } & \colhead{Class}  & \colhead{$\gamma$}  & \colhead{Comment}
}
\startdata

003002 & 19 41 21.99 & 40 12 02.1 & 12.76 & 1.12 & 2 & 32 &  1.04      &  4.48     &  \nodata & \nodata & BLM &  \nodata  &  Rapid rotator  \\ 
004002 & 19 41 21.87 & 40 11 48.6 & 12.84 & 1.22 & 1 & 4 &  3.47      &  0.12     &  0.37     & 92 & SM &  \nodata  &  \nodata  \\ 
005002 & 19 41 14.76 & 40 11 00.8 & 12.88 & 1.15 & 1 & 4 &  0.79      &  0.21     &  0.64     & 86 & SM &  \nodata  &  \nodata  \\ 
006002 & 19 41 15.93 & 40 11 11.5 & 12.94 & 1.08 & 8 & 13 &  1.46      &  0.04     &  5.10     & 95 & BM & 2.190 & SB1 \\ 
008002 & 19 41 22.45 & 40 12 03.4 & 13.01 & 1.10 & 17 & 13 &  3.23      &  0.35     &  3.38     & 93 & SM &  \nodata  &  \nodata  \\ 
009002 & 19 41 21.34 & 40 11 57.2 & 13.01 & 1.13 & 5 & 11 &  2.86      &  0.66     &  4.13 & \nodata & BLM &  \nodata  &  \nodata  \\ 
010002 & 19 41 13.55 & 40 12 20.6 & 13.06 & 1.17 & 1 & 4 &  2.37      &  0.23     &  0.70     & 95 & SM &  \nodata  &  \nodata  \\ 
011002 & 19 41 13.45 & 40 11 56.2 & 13.12 & 1.18 & 5 & 3 &  2.96      &  0.08     &  0.41     & 94 & SM &  \nodata  &  \nodata  \\ 
006003 & 19 41 12.79 & 40 12 23.9 & 13.12 & 1.19 & 1 & 4 &  2.79      &  0.05     &  0.16     & 95 & SM &  \nodata  &  \nodata  \\ 
003005 & 19 41 29.54 & 40 12 21.0 & 13.07 & 1.23 & 6 & 4 &  2.72      &  0.10     &  0.57     & 95 & SM &  \nodata  &  \nodata  \\ 
004005 & 19 41 21.48 & 40 13 57.3 & 13.08 & 1.21 & 15 & 3 &  1.50      &  0.32     &  2.81     & 93 & SM &  \nodata  &  \nodata  \\ 
005005 & 19 41 08.59 & 40 13 29.9 & 13.11 & 1.21 & 1 & 4 &  2.00      &  0.21     &  0.64     & 95 & SM &  \nodata  &  \nodata  \\ 
001006 & 19 41 17.76 & 40 09 15.9 & 12.84 & 1.11 & 4 & 13 &  1.68      &  1.04     &  6.48 & \nodata & BLM &  \nodata  &  \nodata  \\ 
002006 & 19 41 29.15 & 40 13 04.1 & 13.11 & 1.20 & 3 & 0 &  1.01      &  0.11     &  0.39     & 89 & SM &  \nodata  &  \nodata  \\ 
002007 & 19 41 05.24 & 40 14 04.2 & 13.13 & 1.20 & 3 & 4 &  3.25      &  0.18     &  0.74     & 93 & SM &  \nodata  &  \nodata  \\ 
003007 & 19 41 09.26 & 40 14 43.6 & 13.13 & 1.20 & 3 & 4 &  2.30      &  0.28     &  1.17     & 95 & SM &  \nodata  &  \nodata  \\ 
003009 & 19 41 09.91 & 40 15 49.6 & 12.92 & 1.20 & 1 & 5 &  2.36      &  0.20     &  0.68     & 95 & SM &  \nodata  &  \nodata  \\ 
004009 & 19 41 30.27 & 40 15 21.7 & 12.98 & 1.20 & 2 & 3 &  1.77      &  0.24     &  0.80     & 94 & SM &  \nodata  &  \nodata  \\ 
006009 & 19 41 34.44 & 40 08 46.1 & 13.14 & 1.20 & 3 & 2 &  1.28      &  0.28     &  1.02     & 92 & SM &  \nodata  &  \nodata  \\ 
003011 & 19 40 57.97 & 40 08 17.5 & 12.95 & 1.18 & 10 & 7 &  3.97      &  0.63     &  4.67 & \nodata & BLM &  \nodata  &  \nodata  \\ 
004011 & 19 40 50.20 & 40 13 11.0 & 13.09 & 1.18 & 1 & 3 &  3.26      &  0.32     &  0.86     & 93 & SM &  \nodata  &  \nodata  \\ 
002012 & 19 41 16.32 & 40 05 50.9 & 13.11 & 1.20 & 13 & 6 &  3.73      &  0.04     &  6.20     & 95 & BM & 2.718 & SB1 \\ 
008013 & 19 41 02.03 & 40 06 28.1 & 13.14 & 1.14 & 3 & 4 &  2.56      &  0.09     &  0.39     & 95 & SM &  \nodata  &  \nodata  \\ 
003021 & 19 41 23.86 & 40 21 44.5 & 13.02 & 1.14 & 7 & 0 &  1.74      &  0.07     &  0.45     & 94 & SM &  \nodata  &  \nodata  \\

\enddata
\end{deluxetable*}

\subsubsection{Red Clump Stars}

We mark the approximate location of the red clump within the dashed rectangle
(defined as $12.7 \lesssim V \lesssim 13.2$ and $1.06 \lesssim (\bv)
\lesssim 1.23$) in the bottom panel of Figure~\ref{all-mem-cmds}.  In
Table~\ref{rctab} we list 24 potential red clump stars, and provide the
same information as in our full RV database.  Nearly one quarter (6/24) of
the stars in the red clump are velocity variables.  Five of these
binaries likely have long periods (P $>$ 500 days). The sixth, WOCS
\thatbinid, displays properties different from the rest and is worthy of
particular attention.

\paragraph{\thatbinid : }
The red clump binary \thatbinid\ has a circular orbit with a period of
only 17.7 days. Such a short period is not expected for a binary with a
primary star that has evolved through the tip of the giant branch and
back to the red clump.  Indeed, the current orbital separation would not
permit a binary containing a star at the tip of the red giant branch
without significant mass transfer and possibly a common-envelope
phase. 

On the other hand, the circular orbit would not be expected for a main-sequence
non-member binary \citep[e.g.,][]{meibom05}. Thus the circular orbit is
additional evidence supporting \thatbinid\  being an evolved cluster member.

We suggest that the current state of the system may be the result of a dynamical encounter (or encounters). Such an encounter may have exchanged a more massive horizontal-branch primary star into the binary. This large-radius, deeply convective primary star would then rapidly circularize the orbit. 
Our preliminary investigations of this hypothesis show that
this scenario is possible \citep{gosnell07}. We will discuss the
likelihood of such an evolutionary history for \thatbinid\ in detail in
a future paper.

\section{Conclusions}

In this paper we have described our high-precision radial-velocity study
of solar-like stars within 23 arcmin square (13 pc) of the
intermediate-aged open cluster NGC 6819.  We analyzed all available 
photometry and astrometry for the cluster in order to define a sample of 
candidate members, as reported in Table~\ref{t:phot}. We present the current results of our
ongoing comprehensive RV survey of the cluster using the WIYN 3.5m
telescope and the Hydra MOS.  We supplement these RVs with an earlier CfA
data set that, for some stars, significantly extends our time baseline,
allowing us to detect and find orbital solutions for longer period
binaries than would be possible with the WIYN data set alone.  In Table~\ref{the-table} we show the combined RV database, including 
membership and binary status. The result of our work is a complete
sample of all giants, subgiants and blue stragglers in the core of NGC~6819, as well as
identification of a large, though not complete, sample of upper main sequence cluster members. Of these \nummembers\ cluster members, \numvarmem\ appear to be hard binaries. 

In \S~\ref{disc} we use our precise RV membership probabilities to
construct a cleaned CMD of the cluster.  Our analysis confirms the
presence of core convective overshoot in the turnoff stars of the
cluster, as indicated by the now evident morphology at the top of the main
sequence.  We identify a rich population of blue straggler cluster
members, and list their properties in Table~\ref{bstab}.  Additionally,
we list probable red clump members in Table~\ref{rctab}. We also
identify a binary in the red clump that, because of its short period of
only 17.7 days, we conjecture is the product of a dynamical
encounter. 

In this survey we have shown the efficacy of RV measurements for
unbiased surveys of open cluster membership, even when the cluster is
rich, located within the plane of the Galaxy, and has a velocity
distribution that is not entirely distinct from that of the field
population.  With this membership information in hand, NGC 6819 is ripe
for further detailed analysis.

The WIYN Open Cluster Study will continue to study NGC~6819.  In future 
papers, we will investigate the dynamical state of the cluster (e.g., mass segregation and velocity 
distribution) and provide the parameters
for all binaries with orbital solutions, allowing us to study their
distributions as well as to constrain the overall hard-binary frequency of the
cluster.  NGC 6819 is a benchmark intermediate-aged open cluster, and
provides a critical constraint on the evolution of open clusters with
rich binary populations.

\acknowledgments

The authors are grateful to the following individuals for invaluable
contributions to this work: Joanne Rosvick and Jason Kalirai for
providing their photometry data for our comparisons; Bob Davis, Jim
Peters, Perry Berlind, Ed Horine, Joe Zajac and Ale Milone for their
work on the CfA RV observations; Nigel Sharp for taking our original
cluster photometry; Chris Dolan for initial setup work for the survey;
and undergraduate and REU students Nick Stroud, Sylvana Yelda, Meagan
Morscher, Michael DiPompeo and Natalie Gosnell. This research was funded in part by NSF grants
AST-0406615 (R. D. M.) and AST-0406689 (I. P.). K. T. H. gratefully
acknowledges support from the Wisconsin Space Grant Consortium and
Oberlin College.


\bibliographystyle{apj}                       

\bibliography{ngc6819.ms}


\end{document}